\documentclass[preprint,12pt]{article}

\usepackage[hidelinks]{hyperref}
\usepackage{xcolor}

\usepackage{graphicx}
\usepackage{amsmath} 

\usepackage{makecell}
\usepackage{array}
\usepackage{ragged2e}

\begin{document}
\begin{center}
    {\large Investigation of the spatial resolution of PET imaging system measuring polarization-correlated Compton events}\footnote{Manuscript accepted at Nuclear Instruments and Methods in Research Section A. DOI: 10.1016/j.nima.2024.169795.\newline © 2024. This manuscript version is made available under the CC-BY-NC-ND 4.0 license https://creativecommons.org/licenses/by-nc-nd/4.0/}
    \newline
    
    Ana Marija Kožuljević$^{1, *}$, Tomislav Bokulić$^{1}$, Darko Grošev$^{2}$, Zdenka Kuncic$^{3}$, Siddharth Parashari$^{1}$, Luka Pavelić$^{4}$, Mihael Makek$^{1}$
\end{center}

\begin{flushleft}
\textit{{\small $^{1}$University of Zagreb Faculty of Science, Department of Physics, Bijenička cesta 32, 10 000, Zagreb, Croatia}
\newline
{\small $^{2}$Department of Nuclear Medicine and Radiation Protection, University Hospital Centre Zagreb,  Kišpatićeva ulica 12, 10 000, Zagreb, Croatia}
\newline
{\small $^{3}$School of Physics, University of Sydney, Physics Road, 2050, Camperdown, NSW, Australia}
\newline
{\small $^{4}$Institute for Medical Research and Occupational Health, Department of Radiation Protection, Ksaverska cesta 2, 10 000, Zagreb, Croatia}
}
\end{flushleft}
*Corresponding author: amk.phy@pmf.unizg.hr

\begin{abstract}
Recent studies of positron emission tomography (PET) devices have shown that the detection of polarization-correlated annihilation quanta can potentially reduce the background and creation of false lines of response (LORs) leading to improved image quality. We developed a novel PET demonstrator system, capable of measuring correlated gamma photons with single-layer Compton polarimeters to explore the potential of the method. We tested the system using sources with clinically relevant activities at the University Hospital Centre Zagreb. Here we present, for the first time, the images of two Ge-68 line sources, reconstructed solely from the correlated annihilation events. The spatial resolution at two different diameters is determined and compared to the one obtained from events with photoelectric interaction.
\end{abstract}

Keywords: positron emission tomography, Compton polarimeter, polarization correlations, Compton scattering, entanglement

\section{Introduction}
Positron emission tomography (PET) is a medical imaging modality that utilizes $\beta^{+}$ decay of a radiopharmaceutical. The emitted positron ($e^{+}$) annihilates with an electron and in this process, two entangled photons are emitted back-to-back with 511 keV energy and orthogonal polarizations. The mechanism for probing polarization correlations of the annihilation photons is the double Compton scattering, where each photon is scattered with Compton scattering angle $\theta$ and azimuthal scattering angle $\phi$. The scattering cross-section is given by the Pryce-Ward formula \cite{ref-knf}:
\begin{equation}
\frac{\mathrm{d}^{2} \sigma}{\mathrm{d} \Omega_{1} \mathrm{~d} \Omega_{2}} = \frac{r_{e}^{4}}{16} F\left(\theta_{1}\right) F\left(\theta_{2}\right)\left\{1-\frac{G\left(\theta_{1}\right) G\left(\theta_{2}\right)}{F\left(\theta_{1}\right) F\left(\theta_{2}\right)} \cos \left[2\left(\phi_{1}-\phi_{2}\right)\right]\right\}
\label{eqn1}
\end{equation}
with $F\left(\theta_{i}\right) = \frac{\left[2+\left(1-\cos \theta_{i}\right)^{3}\right]}{\left(2-\cos \theta_{i}\right)^{3}}$ and $G\left(\theta_{i}\right) = \frac{\sin ^{2} \theta_{i}}{\left(2-\cos \theta_{i}\right)^{2}}$. The correlations in their polarizations are reflected in the difference of the azimuthal scattering angles $\phi_{1}$ and $\phi_{2}$, such that the highest probability is for $\left|\phi_{1}-\phi_{2}\right| = 90^{\circ}$. The sensitivity of the measurements to this process is given by the polarimetric modulation factor $\mu$:
\begin{equation}
\mu=\frac{G\left(\theta_{1}\right) G\left(\theta_{2}\right)}{F\left(\theta_{1}\right) F\left(\theta_{2}\right)}
\label{eqn2}
\end{equation}
reaching the maximum of 0.48 for ideal detectors and photons scattered with $\theta_{1}=\theta_{2}$ $\approx$ $82^{\circ}$. Measuring the correlated annihilation photons can help distinguish true coincidences from the randoms, since they do not exhibit the same correlation \cite{ref-kuncic, ref-macnamara, ref-toghyani, ref-watts, ref-kim}. To explore this, yet unexploited property in PET, we built a novel PET demonstrator system utilizing single-layer Compton polarimeters \cite{ref-nim, ref-sid2}, capable of measuring the polarization correlations of the annihilation quanta. In this work, we investigate the spatial resolution of the system using two Ge-68 line sources with clinically relevant activities.

\section{The experimental setup}
\label{sec1}
\begin{figure}[ht]
\centering
\includegraphics[width = 10 cm]{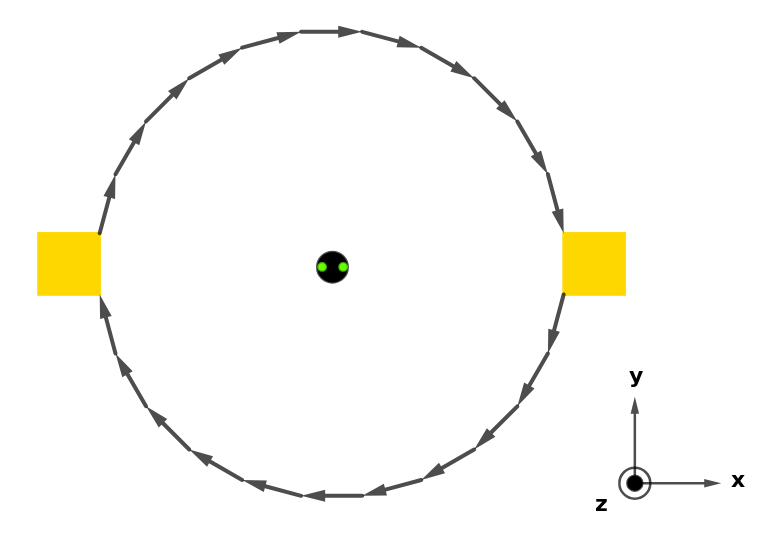}
\caption{Schematic drawing of the axial view of the PET demonstrator (not to scale) at 430 mm diameter (in yellow), with an epoxy phantom 3 cm in diameter (in black) and two line sources (in green). Arrows represent the positions taken by the detectors during data acquisition.}\label{fig:shematics}
\end{figure}

We constructed a PET demonstrator system comprising detector modules based on single-layer Compton polarimeters \cite{ref-nim, ref-crystals, ref-sid2}. The modules consist of 16x16 scintillating crystals, assembled from four 8x8 crystal matrices. Each matrix is coupled to a silicon photomultiplier (SiPM, Hamamatsu Photonics, model S13361-0808AE) and read out by the TOFPET2 acquisition system \cite{ref-tofpet}. Each module is equipped with a dedicated cooling system based on the Peltier unit to dissipate the heat generated by the ASICs and to ensure temperature stability for the SiPMs. 
The utilized crystals are GAGG:Ce crystals with a matrix pitch of 3.2 mm. Individual crystal length is 20 mm, with sides either 2.9 mm x 2.9 mm or 3.0 mm x 3.0 mm \cite{ref-sid2}. The mean energy resolutions are (8.1$\pm$1.1)\% for Module I and (9.3$\pm$2.2)\% for Module II \cite{ref-sid}. The assembled modules are mounted on an aluminum gantry's arms, allowing a diameter range from 420 mm to 700 mm. Four polytetrafluoroethylene (PTFE) rollers support the gantry, and the roller connected to the stepper motor grants precise rotation capability for emulating the full ring of detectors. A sliding wooden platform provides the support and accurate positioning of the sources in the FOV of the scanner for the imaging process. A scheme representing the acquisition setup is shown in Figure \ref{fig:shematics}.

\section{Data acquisition and analysis}
\label{sec2}
\begin{figure}[ht]
\centering
\includegraphics[width = 8 cm]{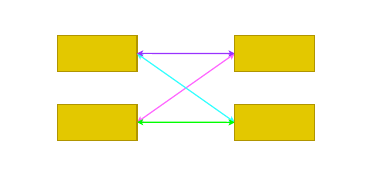}
\caption{Schematic drawing of LOR possibilities in double Compton events. Due to the uncertainty of the scatter pixel determination in each module, four LORs are possible. We select the one connecting the pixels with lower energy on each side.}\label{fig:lor_amb}
\end{figure}

\begin{figure}[ht]
\centering
\includegraphics[width = 8 cm]{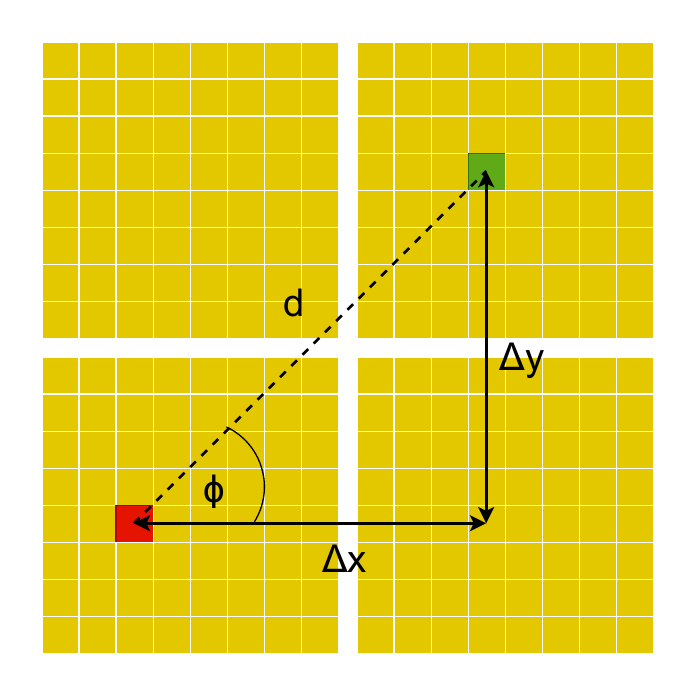}
\caption{Cross-sectional schematics of the detector module. The azimuthal scattering angle $\phi$ is calculated from the Eq. \ref{eqn:comp_kin}, where the red square denotes the pixel with $E_{pix_1}$ (scatterer), and the green square with $E_{pix_2}$ (absorber). The distance between the pixel centers is labeled as $d$.}\label{fig:phi_def}
\end{figure}

Data were obtained at the University Hospital Center Zagreb, where two Ge-68 line sources ($t_{1/2}$ = 270 days) in an aluminum encapsulation, each with activity $\sim$ 45.5 MBq, were placed in a 3 cm thick epoxy phantom, at $\sim2$ cm distance. This arrangement was imaged with scanner's diameter set to 430 mm in one measurement, and at 620 mm in the other. For measurements at the diameter of 430 mm, the detectors were rotated around the scanner axis at 12 positions, with 15$^{\circ}$ steps. In this way, we emulated a ring of detectors with a total of 6144 crystals distributed along 16 axial rings. Measurements at the diameter of 620 mm were taken at 16 positions by rotating the modules in 11.25$^{\circ}$ steps, thus emulating a ring with 8192 pixels. The acquisition length at each position was approximately 2.3 hours for the 430 mm diameter and 1 hour for the 620 mm diameter. The TOFPET2 read-out system was utilized to trigger on coincidence events between the two detector modules when each of the modules fired with at least one pixel. The raw data was first decoded so that hits with  timestamps within 100 ns time window are grouped into one event and written in the ROOT format. Data analysis was performed on a multi-core PC and accelerated with the help of GNU command parallel \cite{ref-parallel}. If, in one such event, only one pixel in the module fired (single-pixel event) with the deposited energy within $\pm$3$\sigma$ of 511 keV, it is labeled a "photoelectric event". If two pixels fired in each module, and the sum of deposited energies is within $\pm$3$\sigma$ of 511 keV with energies relevant to Compton scatterings kinematics, the event is recognized as double Compton scattering. It is not always possible to deduce which of the two pixels in the Compton scattering interaction fired first (the scatterer) due the similar energy deposits in both pixels. However, the detector geometry and the interaction's cross-section favor forward scattering, hence we select the pixel with the lower deposited energy as the scatterer ($E_{pix_1}<E_{pix_2}$). Simulation studies show this selection is true for \~55\% of cases\cite{ref-amk}. This introduces an uncertainty in the line-of-response (LOR) determination (Fig. \ref{fig:lor_amb}), which affects the spatial resolution.  
After the pixels are selected and their order of firing determined, the Compton scattering angles are reconstructed as:
\begin{eqnarray}
\theta  =  \text{acos}\left(\text{m}_{e}\text{c}^{2}\left(\frac{1}{E_{pix_1}}-\frac{1}{E_{pix_2}}\right)+1\right), \phi= \text{atan}\left(\frac{\Delta y}{\Delta x}\right)
\label{eqn:comp_kin}
\end{eqnarray}
where the azimuthal scattering angle is determined from the geometry, as can be seen in Figure \ref{fig:phi_def}.
The detectors have non-uniform azimuthal acceptance since the combinations of the two pixels that fired in a Compton event depend on the distance $d$ between the scatterer and the absorber. With increasing $d$, the probability of observing an event is reduced. Therefore the distribution of the difference of the obtained angles N($\phi_1-\phi_2$) is corrected:
\begin{equation}
N_{\text {cor}}\left(\phi_{1}-\phi_{2}\right) = \frac{N\left(\phi_{1}-\phi_{2}\right)}{A_{mix}\left(\phi_{1}-\phi_{2}\right)},
\label{eq:accept_corr}
\end{equation}
where $A_{mix}(\phi_{1}-\phi_{2})$ is the $\phi_{1}-\phi_{2}$ distribution obtained by event mixing technique. With this technique, the azimuthal difference is obtained from physically different, thus uncorrelated events \cite{ref-nim}.  The polarimetric modulation factor $\mu$ is then determined by fitting the relation $N_{\text {cor }}\left(\phi_{1}-\phi_{2}\right) = M\left[1-\mu \cos (2\left(\phi_{1}-\phi_{2}\right))\right]$. The obtained value at 430 mm diameter was $\mu=0.22\pm0.01$, and $\mu=0.20\pm0.01$ at 620 mm diameter.
\begin{figure}[ht]
\centering
\includegraphics[width = 13 cm]{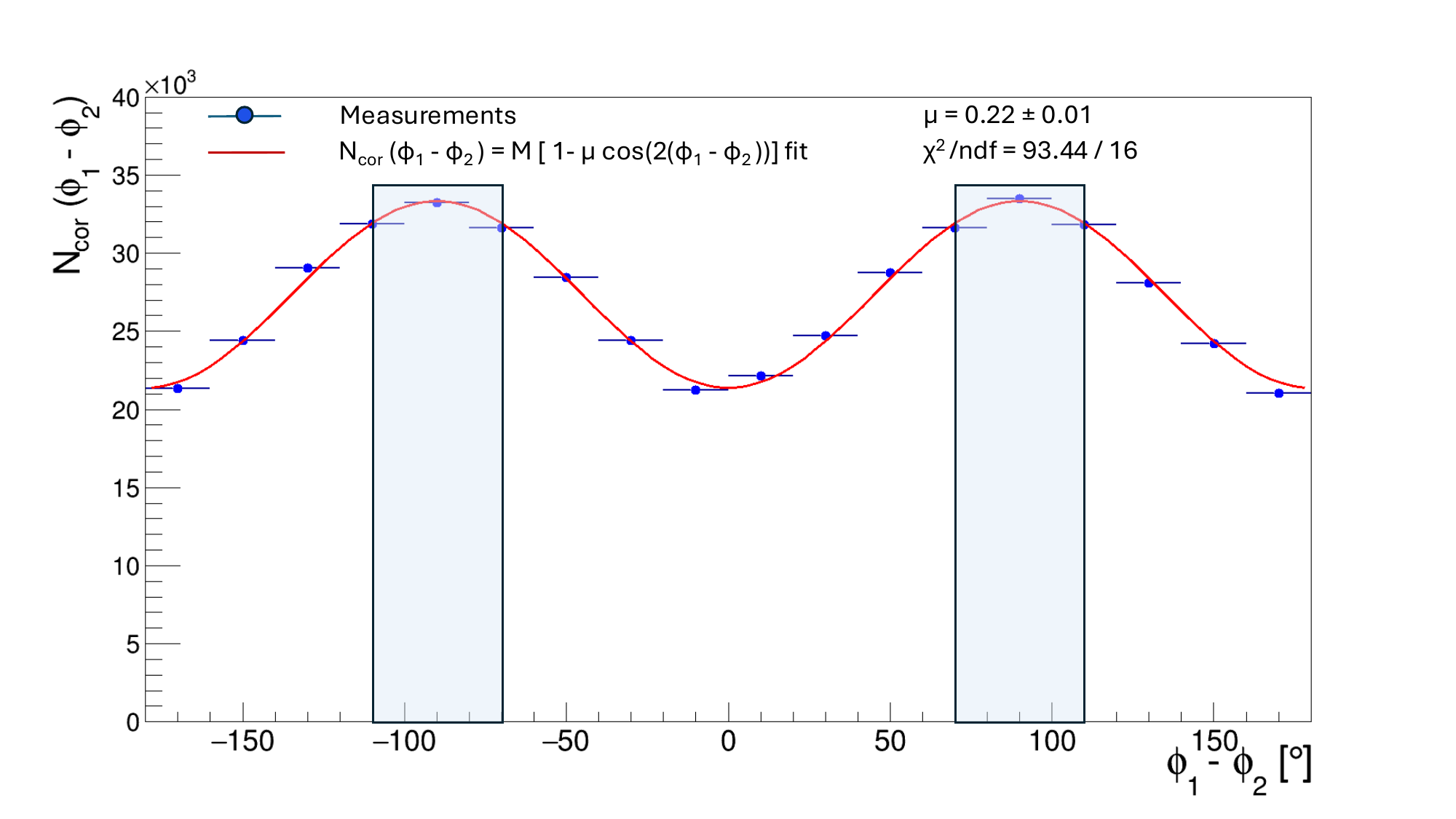}
\caption{Acceptance corrected $\phi_{1}-\phi_{2}$ distribution at 430 mm ring diameter. The Compton scattering angle range was selected as $72^{\circ}<\theta_{1,2}<90^{\circ}$. The highlighted regions denote the azimuthal difference range of events utilized for image reconstruction.}\label{fig:modulation}
\end{figure}
\section{Image reconstruction and analysis}
\label{sec3}
Image reconstruction is done with OMEGA software in MATLAB environment \cite{ref-omega}. The data input is prepared in a list mode so that each event represents one line in the list, with the information on the pixel's energies and their coordinates. To adapt the data for OMEGA reconstruction, measurements at different positions during the acquisition are distributed as if the data were collected with a full ring of detectors. The LORs are then formed between the coordinates of the centers of the listed pixels' faces. In the case of Compton scattering events, the pixels with lower energies ($E_{pix_1}$) are chosen for LOR formation. The \textit{Ordered Subset Expectation Maximization} (OSEM) algorithm is used for the image reconstruction. The FOV is 20x20 cm$^{2}$ and the number of slices is set to 31, determined as the 2n-1, n being the number of axial rings (16). The number of iterations is set to 10, while the number of subsets is varied from minimum of 4 to maximum of 12 or 16, corresponding to data-taking positions for the 430 and 620 ring diameter measurements, respectively. The final images are obtained by summing the image slices to obtain a higher signal-to-noise ratio (SNR). For the axial view, the sum is in the z-direction (31 slices), for the coronal view, the sum is in the y-direction (128 slices), and the sagittal view from the sum of slices in the x-direction (128 slices) (for the reference frame, see Figure 1). The intensity profiles are obtained by summation of the pixel intensities in the coronal direction in the summed axial images and axial direction in the summed coronal images. The intensity profiles are fitted with a double Gaussian function to determine the full width at half maximum (FWHM) of each peak.

\section{Results}
\label{sec4}
\begin{figure}[ht]
\centering
\includegraphics[width = 13 cm]{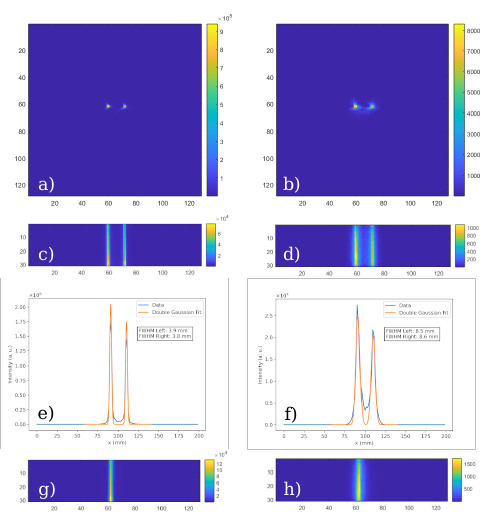}
\caption{Reconstructed images (sums of all slices) of Ga-68 line sources in the phantom, obtained with PET ring diameter of 430 mm. Panels a) and b) represent the axial views, c) and d) are the coronal views, with the corresponding profiles in e) and f), and g) and h) the sagittal views, reconstructed from photoelectric events (left) and using only polarization-correlated events (right), respectively. Images are reconstructed with 10 iterations and 6 subsets.}\label{fig:1800_plots}
\end{figure}

\begin{figure}[ht]
\centering
\includegraphics[width = 13 cm]{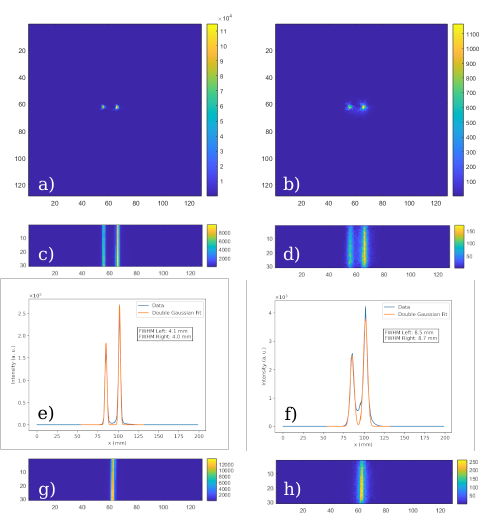}
\caption{Reconstructed images (sums of all slices) of Ga-68 line sources in the phantom, obtained with PET ring diameter of 620 mm. Panels a) and b) represent the axial views, c) and d) are the coronal views, with the corresponding profiles in e) and f), and g) and h) the sagittal views, reconstructed from photoelectric events (left) and using only polarization-correlated events (right), respectively. Images are reconstructed with 10 iterations and 8 subsets}.\label{fig:1794_plots}
\end{figure}

Images of the sources are reconstructed by selecting the events with $\phi_1-\phi_2$ = 90$^{\circ}\pm$20$^{\circ}$, as marked in Figure \ref{fig:modulation}, that happened within a coincidence time window of 5 ns. They are shown in Figures \ref{fig:1800_plots} and \ref{fig:1794_plots} for 430 mm and 620 mm diameters, respectively. No corrections were applied during or after reconstruction. They demonstrate it is possible to reconstruct images by using exclusively correlated annihilation gamma photons, obtained from sources with clinically relevant activities. 

In Table \ref{tab:table} are FWHM values obtained by fitting a double Gaussian function to intensity profiles obtained from the reconstructed images (as in Fig. \ref{fig:1800_plots} and \ref{fig:1794_plots}). There is no substantial difference in the resolutions obtained by increasing the number of subsets in the image reconstruction process.

The spatial resolution of images obtained with single-pixel events, dominated by the photoelectric absorption is superior and reaches approximately 4 mm (FWHM), which is comparable to the state-of-the-art clinical devices. The resolution obtained with polarization-correlated events is approximately 9 mm (FWHM). This degradation is caused by the uncertainty in the LOR creation in two-pixel Compton events, as depicted in \ref{fig:lor_amb}.

\begin{table}[ht]
    \centering
    \small
    \begin{tabular}{|c|c|c|c|}
        \hline
        \multicolumn{4}{|c|}{\textbf{FWHM [mm]}} \\ \hline
        \makecell[c]{\textbf{Ring} \\ \textbf{diameter}}
         & \textbf{Source} & \makecell{\textbf{Photoelectric} \\ \textbf{interaction}} & \makecell{\textbf{Polarization} \\ \textbf{correlations}} \\ \hline
        430 mm & Left & 3.9 & 8.6 \\ \cline{2-4}
         & Right & 3.8 & 8.5 \\ \hline
        620 mm & Left & 4.1 & 8.5 \\ \cline{2-4}
         & Right & 4.0 & 8.7 \\ \hline
    \end{tabular}
    \caption{FWHM obtained for both photoelectric events and events with correlation in polarization at 430 mm and 620 mm diameters. The values presented are obtained from images in Figures \ref{fig:1800_plots} and \ref{fig:1794_plots}, reconstructed with 10 iterations, and 6 subsets for images at 430 mm and 8 subsets for images at 620 mm.
    }
    \label{tab:table}
\end{table}

\section{Discussion and conclusions}
\label{sec5}
We developed the PET demonstrator system utilizing single-layer Compton polarimeters, capable of reconstructing polarization correlations of the annihilation photons. It was successfully tested with two Ge-68 line sources, with a total activity of ~90 MBq at two PET ring diameters, 430 mm and 620 mm. We demonstrated it is possible to reconstruct the source image using only the LORs identified by correlated annihilation quanta. 
The spatial resolution of these images is not considerably influenced by the reconstruction parameters, and the uncertainty in LOR determination in Compton events remains the limiting factor.

\section*{Acknowledgments}
This work was supported in part by Croatian Science Foundation under the project IP-2022-10-3878 and in part by the “Research Cooperability” Program of the Croatian Science Foundation, funded by the European Union from the European Social Fund under the Operational Programme Efficient Human Resources 2014–2020, grant number PZS-2019-02-5829.

\bibliographystyle{unsrt} 
\bibliography{manuscript}

\end{document}